\documentclass{aa} 

\usepackage{graphicx}

\usepackage{array}
\usepackage{lscape}                                
\usepackage{natbib}
\usepackage{longtable}
\usepackage{color}
\usepackage{mathrsfs} 
\usepackage{subcaption}

\usepackage{txfonts}
\newcommand{\kms} {km\,s$^{-1}$}
\newcommand{\vsini} {$v$\,sin\,$i$}
\newcommand{\vmac} {$v_ {\rm mac}$}
\newcommand{\Teff} {T$_{\rm eff}$}
\newcommand{\grav} {log\,{\em g}}

\newcommand{\rvpp}{RV$_{\rm pp}$}

\begin{document} 

   \title{The IACOB project}

   \subtitle{CVIII. Hunting for spectroscopic binaries in the O and B supergiant domain.\\ The threat of pulsational variability \\ {\em \small (Yet unpublished - but hopefully not LOST - version)}}

   \author{S.~Sim\'on-D\'iaz\inst{1,2}, N.~Britavskiy\inst{3,4}, N. Castro\inst{5,6}, G. Holgado\inst{1,2,7}, A. de Burgos\inst{1,2}}

\institute{
Instituto de Astrof\'isica de Canarias, E-38200 La Laguna, Tenerife, Spain              
\and
Departamento de Astrof\'isica, Universidad de La Laguna, E-38205 La Laguna, Tenerife, Spain
\and 
Royal Observatory of Belgium, Avenue Circulaire/Ringlaan 3, B-1180 Brussels, Belgium
\and
University of Li\`ege, All\'ee du 6 Ao\^ut 19c (B5C), B-4000 Sart Tilman, Li\`ege, Belgium
\and
Institut für Astrophysik, Georg-August-Universität,
Friedrich-Hund-Platz 1, 37077 Göttingen, Germany
\and
Leibniz-Institut für Astrophysik Potsdam (AIP), An der Sternwarte 16,
14482, Potsdam, Germany 
\and 
Centro de Astrobiolog\'ia (CAB), CSIC-INTA, Camino Bajo del Castillo s/n, campus ESAC, 28692, Villanueva de la Cañada, Madrid, Spain
}

\offprints{ssimon@iac.es}

\date{Submitted/Accepted}

\titlerunning{Hunting for spectroscopic binaries in Galactic OB-Sgs}

\authorrunning{Sim\'on-D\'iaz et al.}

 
  \abstract
   {Observations have definitively strengthened the long-standing assertion that binaries are crucial in massive star evolution. While the percentage of spectroscopic binary systems among main-sequence O stars is well-studied, other phases of massive star evolution remain less explored.}
   {We aim to estimate the spectroscopic binary fraction in Galactic late O- and B-type supergiants (OB-Sgs) and set empirical thresholds in radial velocity (RV) to avoid misidentifying pulsating stars as single-line spectroscopic binaries.}
   {Using over 4500 high-resolution spectra of 56 Galactic OB-Sgs (plus 13 O dwarfs/subgiants and 5 early-B giants) from the IACOB project (2008-2020), we apply Gaussian fitting and centroid computation techniques to measure RV for each spectrum.}
   {Our findings reveal that intrinsic variability in OB-Sgs can result in peak-to-peak RV amplitudes (\rvpp) of up to 20\,-\,25~\kms, notably in late-O and early-B Sgs, and decreases to typical values of \rvpp\ in the range of 1\,-\,5~\kms\ for O dwarfs and 2\,-\,15~\kms\ for late B-Sgs. Considering these results and evaluating line-profile variability in each star, we find that 10$\pm$4\% of OB-Sgs are clearly single-line spectroscopic binaries. In addition, we find that the percentage of double-line spectroscopic binaries in late O- and early B-Sgs is ~6\%, much lower than the ~30\% in O-type dwarfs and giants. 
   }
   {This study, along with prior research on B-Sgs in the 30 Doradus region of the LMC, indicates that the spectroscopic binary percentage decreases by a factor of 4\,-\,5 from O stars to B-Sgs. Our study also underscores the need for a thorough characterization of spectroscopic variability due to intrinsic sources to reliably determine the spectroscopic binary fraction among OB-Sgs and O-type stars in general, offering valuable insights into the impact of binaries on massive star evolution.}


   \maketitle

\section{Introduction}\label{intro}

Recent spectroscopic and high-angular resolution surveys of O-type stars have provided fresh insights on the multiplicity properties of main sequence massive stars \citep[][and references therein]{Sana2017, Barba2017, Vanbeveren2017}. On one hand, the observed fraction of spectroscopic binary/multiple systems with at least one O star component reported by several medium and large size spectroscopic surveys ranges between 42\% and 68\% \citep[see, e.g.,][]{Barba2010, Sana2012, Sana2013, Chini2012, Sota2014, Kobulnicky2014, Aldoretta2015, Almeida2017}. On the other hand, as indicated by \cite{Sana2014}, when information from high-angular resolution surveys is also considered, the observed multiplicity fraction of Galactic O-type stars with close-by companions can reach 90\%.

These observational results have certainly reinforced the long-standing assertion that binary interaction should not be overlooked when investigating massive star evolution \citep[e.g.,][]{Vanbeveren1993}. Indeed, based on empirical arguments, \cite{Sana2012} proposed that 25\% of massive binary systems will interact while both components are still on the main sequence, and nearly 70

In parallel, the second decade of the 21st century witnessed the first direct observation of gravitational waves originating from a pair of merging black holes \citep{Abbott2016}. Since then, several tens of detections have been announced, including several cases associated with the merging of two neutron stars \citep{Abbott2017} and a potential event produced by a neutron star-black hole binary merger \citep{Ackley2020}.  

Theoreticians and modelers are working diligently to investigate the impact that binary interaction may have on our current understanding of massive star evolution \citep[e.g.,][]{Song2013, Song2016, Song2018, deMink2013, deMink2016, Eldridge2017, Wang2020}. They are also studying the various possible pathways toward the merging of massive black holes and neutron stars \citep[e.g.,][]{Marchant2016, Marchant2017, Eldridge2016, Eldridge2019, Mandel2016, deMink2016, Stevenson2017}. 

Slowly but surely, we are also increasing the amount of available information about the orbital and physical properties of O-type binaries \citep[e.g.,][]{Almeida2017, Barba2017, Sana2017, Martins2017, MaizApellaniz2019, Trigueros2021, Mahy2022, Shenar2022}. This is allowing us to construct empirical distributions of binary periods, mass ratios, and eccentricities using statistically significant samples of binary and multiple systems in different environments. Additionally, we can identify hidden companions spectroscopically, which could be associated with the donor of a post-interaction binary system after becoming a stripped star, a black hole, or a neutron star. 

The time is ripe to conduct similar observational studies in other phases of the evolution of massive stars beyond the main sequence, such as the so-called blue and red supergiants. This step is crucial and necessary if we aim to empirically constrain any theoretical attempt to connect the early evolutionary phases of massive binaries (i.e., the O and B main sequence stars) with their potential end products, represented by black hole and neutron star binaries that lead to gravitational wave emission if they merge at some point in their lifetimes. 

In this paper, we primarily focus on the more direct descendants, from an evolutionary standpoint, of main sequence O-type stars: the B-type supergiants. Despite their importance, the multiplicity properties of this domain of massive star evolution have been only vaguely explored to date \citep[e.g.,][]{Dunstall2015, McEvoy2015}. We present results from the multi-epoch spectroscopic analysis of a sample of 61 evolved O- and B-type stars (mostly supergiants, but also a few giants and bright giants) and 13 O-type subgiants/dwarfs for which we have more than 4500 high-resolution spectra gathered over a period of about 12 years. As we will illustrate throughout this paper, the ubiquitous existence of pulsational-type phenomena in this region of the Hertzsprung-Russell diagram makes the correct identification of spectroscopic binaries in O- and B-type supergiants (OB-Sgs) more challenging than in O giants (Gs) and dwarfs (DWs). In this case, a more complex observational strategy is required to avoid spurious conclusions about the percentage of binaries resulting from the misidentification of pulsating stars as single-line spectroscopic binary (SB1) systems.   

The paper is structured as follows. In Sect.~\ref{section2}, we define the working sample and present the observational material. The tools and methods used to extract information about the stellar parameters and perform the radial velocity (RV) measurements are presented in Sect.~\ref{section3}. Results of our investigation of the percentage of spectroscopic binaries in the B-Sg domain are presented in Sect.~\ref{section4} and discussed in Sect.~\ref{section5}. The main conclusions and some guidelines for future investigations are summarized in Sect.~\ref{summary}.

\section{Observations and sample definition}\label{section2}

The bulk of the observations considered in this paper were gathered within the framework of the IACOB project \citep{SimonDiaz2011b, SimonDiaz2015, SimonDiaz2020}. This includes multi-epoch, high-resolution spectra obtained with the FIES instrument \citep{Telting2014} attached to the 2.56-meter Nordic Optical Telescope (NOT), and the HERMES spectrograph \citep{Raskin2011} attached to the 1.2-meter Mercator telescope during several observing runs allocated between 2008 and 2018. As we have indicated in other papers of the IACOB series, although HERMES provides a somewhat larger resolving power ($R$\,=\,85\,000) than the mid- and high-resolution mode of FIES observations ($R$\,=\,46\,000 and 67\,000, respectively), the combined use of both instruments has allowed us to build an almost homogeneous spectroscopic data set regarding the quality of the spectra, long-term stability of the wavelength calibration, and accuracy of the zero point in radial velocity. Generally speaking, these observations allow us to achieve accuracies in RV measurements ranging from several hundred m\,s$^{-1}$ to a few \kms, mainly depending on the broadening of the diagnostic lines.

\begin{figure}[!t]
\centering
\includegraphics[angle=0, width=0.50\textwidth]{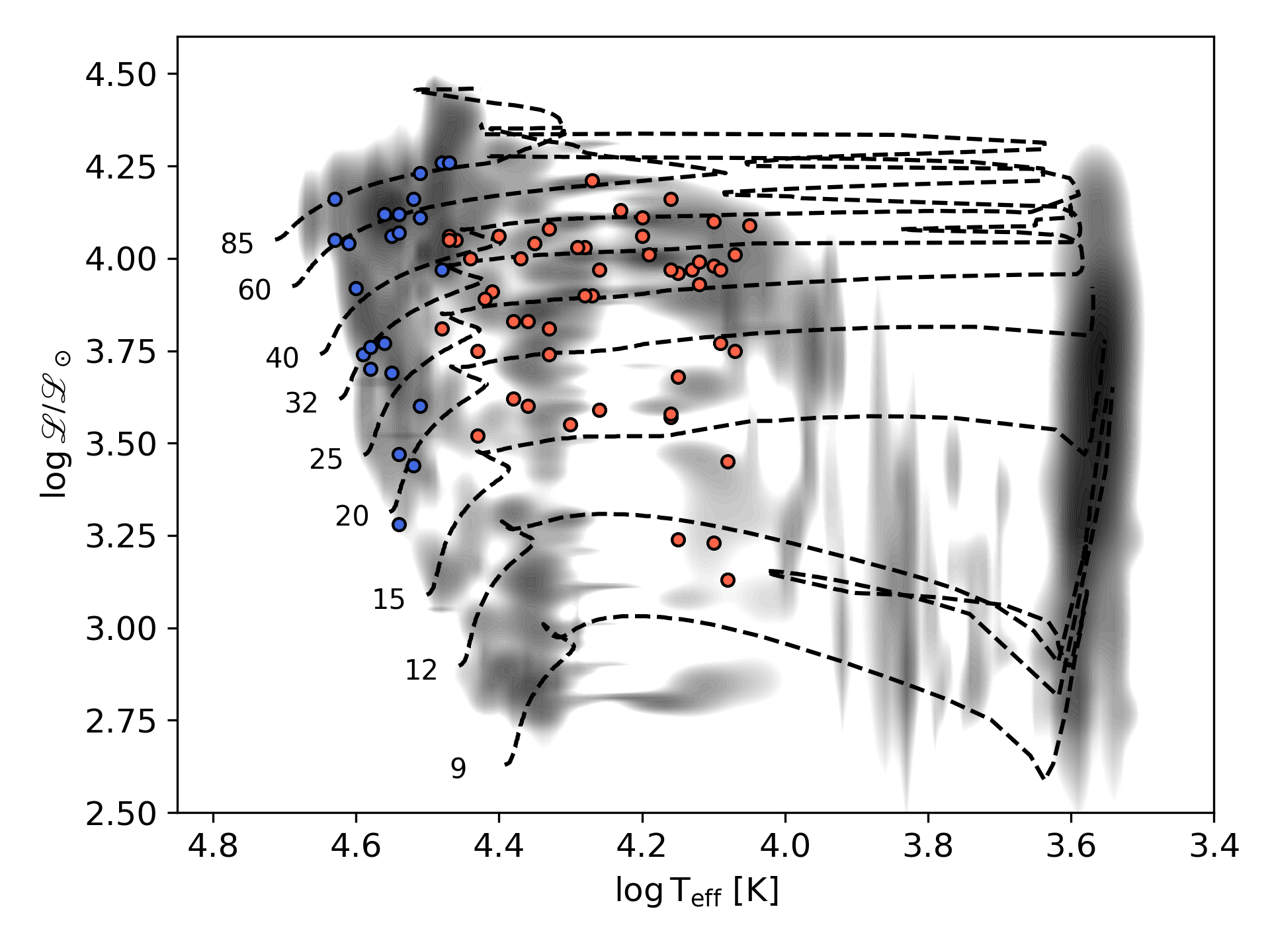}
\caption{Location of the investigated sample of Galactic O stars (blue) and B-Gs/Sgs (red) in the spectroscopic HR diagram. For reference, the figure also depicts in the background a grayscale representation of the probability density distribution of the location of a sample of 575 Galactic massive stars \citep[see][for further details]{Castro2014}.}
\label{figure1}
\end{figure}

By the time the study presented in this paper was performed (late in 2020), the IACOB spectroscopic database\footnote{http://research.iac.es/proyecto/iacob/iacobcat/} comprised more than 8500 spectra of about 1000 Northern Galactic O- and B-type stars \citep[see a more detailed description of the IACOB database and the main drivers for the compilation of such a huge spectroscopic dataset in][]{SimonDiaz2015, SimonDiaz2020}. In this study, we mainly focus on the sample of O-type stars, B supergiants (B-Sgs), and early B giants (early B-Gs) for which we had more than four epochs available. However, we also make use of the complete dataset to identify double-line spectroscopic binaries (SB2), which are detectable in some cases even with access to only one or two epochs (see Sect.~\ref{section4}).

Table~\ref{table:B1} lists our working sample of stars.  This sample comprises 13 O-type dwarfs and subgiants (LCs V and IV), 10 mid/late O-type giants and supergiants (LCs III, II and I), 46 B-Sgs (LCs I and II) and 5 early~B-Gs (LC III). 

Figure~\ref{figure1} depicts the location of these 74 stars in a spectroscopic Hertzsprung-Russell diagram \citep[sHRD,][]{Langer2014}. To serve as a reference, we also present in the background a grayscale representation of the probability density distribution of the location of 575 Galactic massive stars in this diagram, following the method described by \cite{Castro2014}. As expected, our sample of stars mainly covers the hot evolutionary stages of stars with masses in the range of approximately 15--80 M$_{\odot}$, following the single-star non-rotating evolutionary tracks from \cite{Ekstrom2012}.

\begin{figure}[!t]
\centering
\includegraphics[angle=0, width=0.47\textwidth]{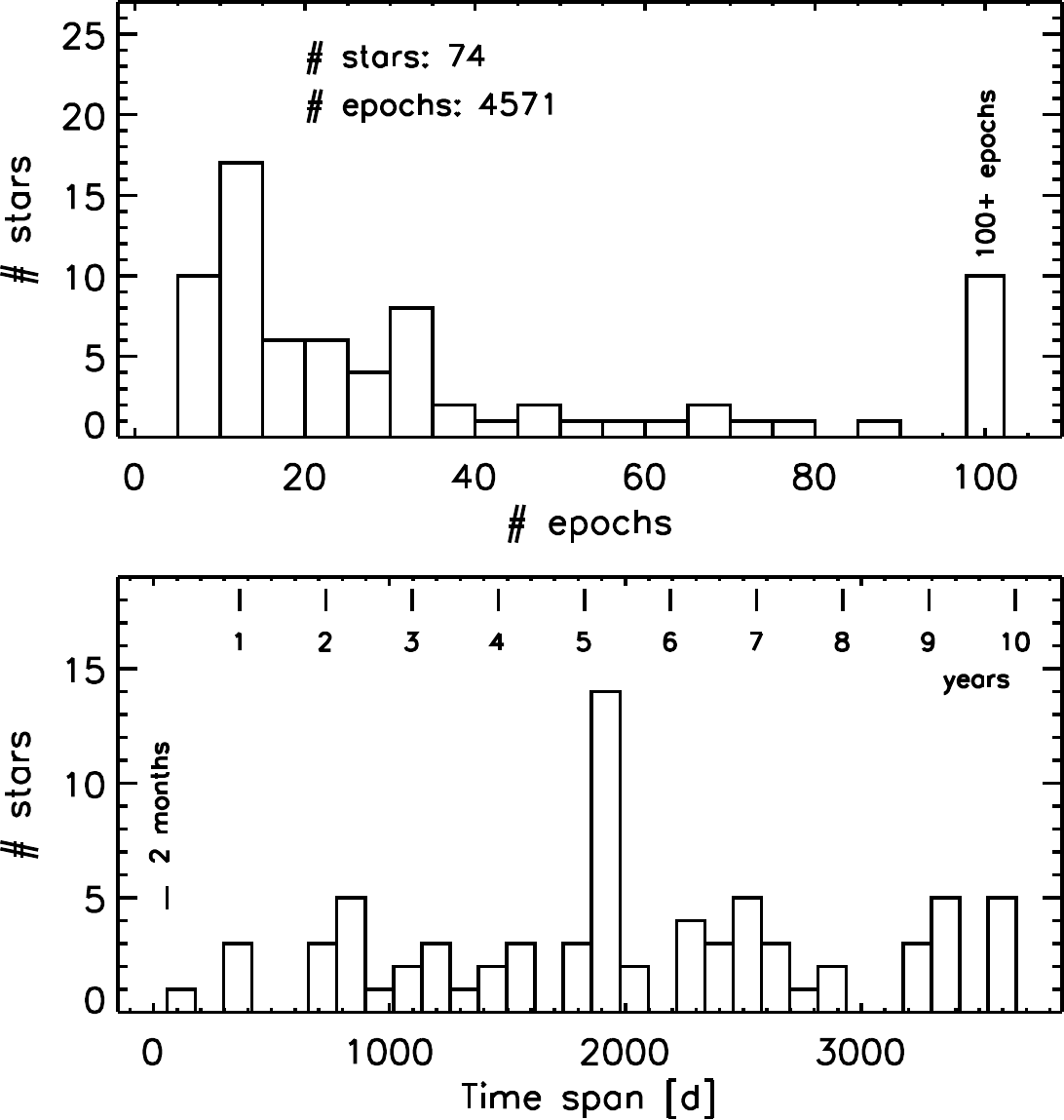}
\caption{Histogram distribution of the number of epochs {\em (top panel)} and the time~span of the observations {\em (bottom panel)} associated with the studied sample of O stars, B-Sgs, and early~B-Gs not identified as SB2.}
\label{figure2}
\end{figure}

Figure~\ref{figure2} shows two histograms indicating the number of stars with a certain number of epochs (top) and the characteristic time-span of the compiled observations (bottom). As seen in the top panel of Fig.~\ref{figure2}, we have access to 5\,--\,25 spectra for approximately $\sim$60\% of the stars. This is the usual number of epochs considered in most previous works in the literature hunting for spectroscopic binaries in the OB star domain \citep[e.g.,][]{Chini2012, Sana2013, Dunstall2015}. However, there is also a non-negligible number of stars with a significantly larger number of epochs available, with some reaching more than one hundred.

Figure~\ref{figure1a} (complemented with the bottom panel of Fig.~\ref{figure2}) provides a global overview of the time coverage of the observations. As a natural consequence of the way the IACOB spectroscopic database has been built\footnote{By the time of writing this paper, the database included observations gathered during more than 120 nights distributed between 2008 and 2020. Some of them are associated with observing runs devoted to the investigation of spectroscopic signatures of stellar oscillations in O stars and B-Sgs \citep[e.g.,][]{SimonDiaz2010, SimonDiaz2017, SimonDiaz2018, Aerts2017, Aerts2018}.}, our observations are far from being homogeneous both in terms of total time-span and duty-cycle. However, this must not be considered as a caveat for the study presented here, especially considering that for all the stars we have a minimum time span of approximately 2 months and a minimum of 5 epochs.

As described elsewhere \citep[see, e.g.,][]{SimonDiaz2014, Holgado2018}, all spectra were normalized and corrected for Earth's motion in a homogeneous manner using our own routines implemented in IDL and the information provided in the headers of the FITS files. In all cases, we used the interstellar Na~{\sc i} lines at 5890 and 5895 \AA\ as a sanity check of the stability of the zero point in radial velocity for all the spectra of a given target (which we found to be better than 1~\kms).  

In addition to the main HERMES and FIES spectroscopic datasets described above, in this paper, we also benefit from additional time-series spectroscopy  (with a resolution of approximately $R\approx80,000$) obtained with the \emph{Hertzsprung}-SONG 1m telescope \citep{Grundahl2007, Grundahl2014, Fredslund2019}. These observations have been gathered in the framework of a spin-off project of IACOB, in which we are using this robotic telescope to obtain long-term and high-cadence observations of a sample of bright northern O stars and B-Sgs. A specific description of this higher cadence spectroscopic dataset, as well as its scientific exploitation, will be presented in subsequent papers \citep[see also][]{SimonDiaz2017}. In the meanwhile, for the sake of the discussion presented in Sect.~\ref{section52}, we have used some of the gathered time-series. In particular, they allow us to provide some illustrative examples of the type of line-profile variability detected in several presumably single pulsating OB giants and supergiants.

\section{Tools and methods}\label{section3}

\subsection{Spectroscopic parameters}\label{section31}

\begin{figure*}[!t]
\centering
\includegraphics[angle=90, width=0.99\textwidth]{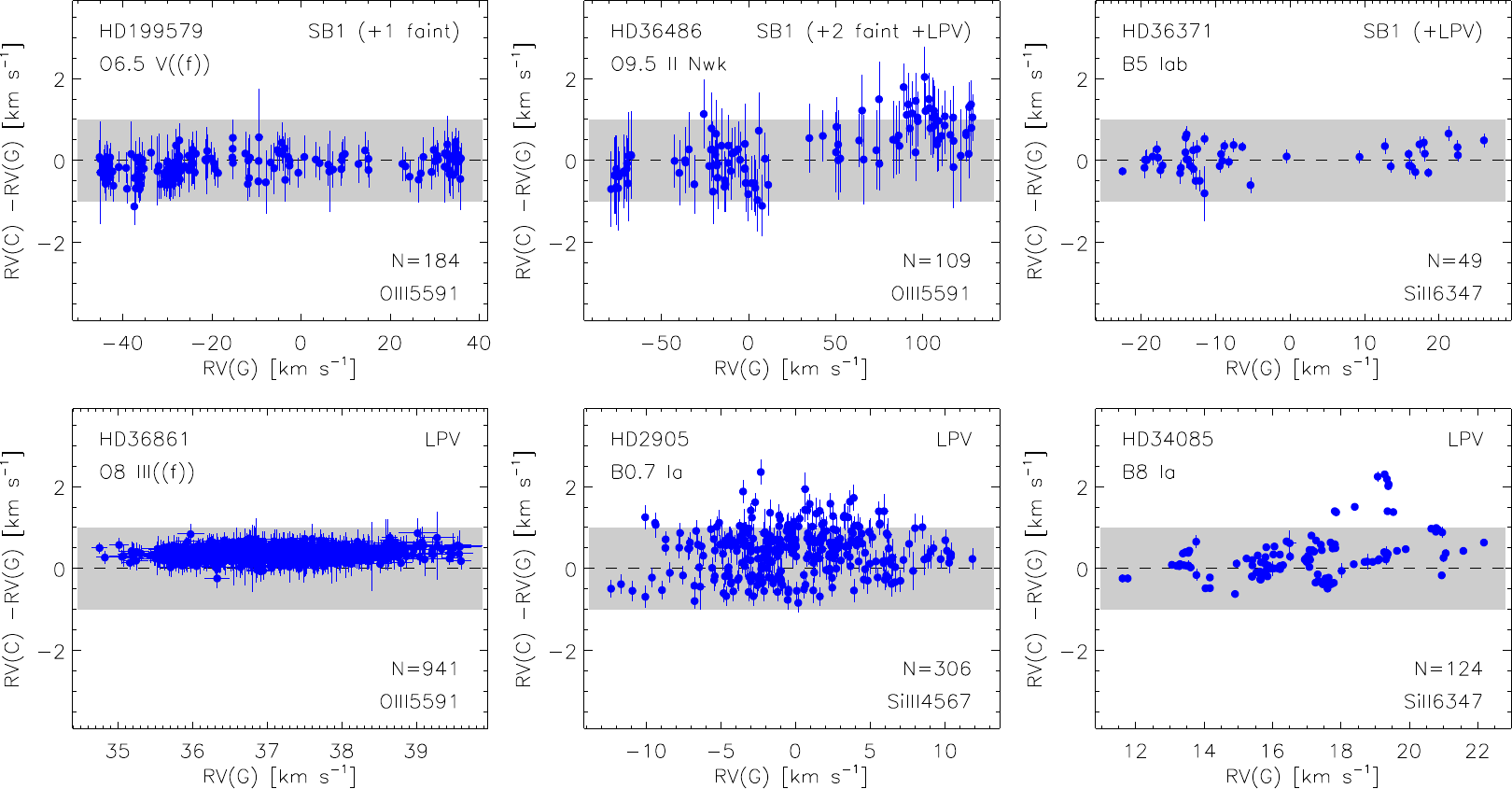}
\caption{measurements obtained by means of Gaussian fitting (G) and computation of the centroid of the line profile (C). The top panels correspond to three clearly detected spectroscopic binaries, while the other three panels present cases in which the detected RV is originated by stellar oscillations (hence labeled as line-profile variable, LPV). The grey horizontal band indicates, for reference purposes, the $\pm$1~\kms\ deviation between the two measured quantities. We also specify at the bottom of each panel the number of available observations and the diagnostic line used for the RV measurements. Note: Although indicated as SB1—since we only detect 1 component in the investigated diagnostic line—\citet{Williams2001} and \citet{Shenar2015} have reported the spectroscopic detection of 1 and 2 faint cooler companions in HD~199579 and HD~36486, respectively.}
\label{rv_cm1}
\end{figure*}

We applied semi-automated techniques to obtain estimates for the projected rotational velocity (\vsini), the amount of macroturbulent broadening (\vmac), the effective temperature (\Teff), and the gravity (\grav) of each star ( in our working sample.

First, we derived the line-broadening parameters (\vsini\ and \vmac) by applying the \textsc{iacob-broad} tool to one of the following diagnostic lines in the best signal-to-noise ratio (S/N) spectrum of each star:  O{\sc iii}~$\lambda$5592, Si{\sc iii}~$\lambda$4567, and Si{\sc ii}~$\lambda$6347 (for the O, early~B, and mid/late~B-type stars, respectively). The resulting values are quoted in columns 9 and 10 of Table~\ref{table:B1}. For a detailed description of the methodology, refer to \cite{SimonDiaz2007, SimonDiaz2014}, and \cite{SimonDiaz2017}.

Next, we proceeded to the quantitative spectroscopic analysis of the O and B star samples using two different but complementary approaches. The O stars were analyzed using \textsc{iacob-gbat} \citep{SimonDiaz2011a}, a grid-based semi-automated tool that automates traditional by-eye techniques based on the analysis of H{\sc i} and He{\sc i-ii} diagnostic lines. We refer the reader to \cite{Holgado2019}, \cite{Holgado2018}, and \cite{Holgado2020} for an in-depth discussion on how these analyses were performed. While \textsc{iacob-gbat} provides estimates (and associated uncertainties) for up to six spectroscopic parameters, we only quote the estimates obtained for \Teff\ and log~$\mathcal{L}$ in columns 7 and 8 of Table~\ref{table:B1}, respectively\footnote{log~$\mathcal{L}$\,=\,4~log\Teff\,-\,\grav\ \citep{Langer2014}.}.

The B-Sg sample was analyzed using an equivalent automatic $\chi^2$ line-profile fitting algorithm following the methodology described in \cite{Castro2012} \citep[see also][]{Lefever2007}. The corresponding quantitative analyses required a dedicated grid of \textsc{fastwind} models comprising the range of effective temperatures and gravities covered by the B-Sgs (and early B-Gs). In this case, in addition to H~{\sc i} and He~{\sc i-ii}, a suitable set of Si~{\sc ii-iii-iv} diagnostic lines were also included in the synthetic spectra to be able to constrain the effective temperature \citep[e.g.,][]{McErlean1999}. A description of the range of stellar parameters covered by this grid and the diagnostic lines used in the analysis can be found in \cite{Castro2012}. Similarly to the case of the O star sample, we only quote the \Teff\ and log~$\mathcal{L}$ estimates resulting from the automated analysis in columns 7 and 8 of Table~\ref{table:B1}.

\subsection{Radial velocity measurements}\label{section32}

We initially considered two different techniques for the measurement of the radial velocity (RV) associated with each individual spectrum of a given star. The first one is based on Gaussian fitting of a selected diagnostic line, providing the quantity that we refer to as RV(G). The second one relies on the computation of the centroid\footnote{This quantity is also known as the first normalized moment in the field of asteroseismology \citep[see, e.g.,][]{Aerts1992}.} of the corresponding line profile, defined as:
\begin{equation}
\mbox{RV(C)}=c\left(\frac{\sum_i \lambda_i(1-f_i)}{\sum_i\lambda_0(1-f_i)}-1\right),
\end{equation}
where, $c$  represents the speed of light in vacuum, $\lambda_{0}$ denotes the central wavelength of the given spectral line, and $\lambda_{i}$ and $f_{i}$ represent the local wavelengths and fluxes of the line profile, respectively. The spectral window used for this computation is defined by the part of the line profile with a local (normalized) flux below 0.98.

\begin{figure*}[!t]
\centering
\includegraphics[angle=90, width=0.99\textwidth]{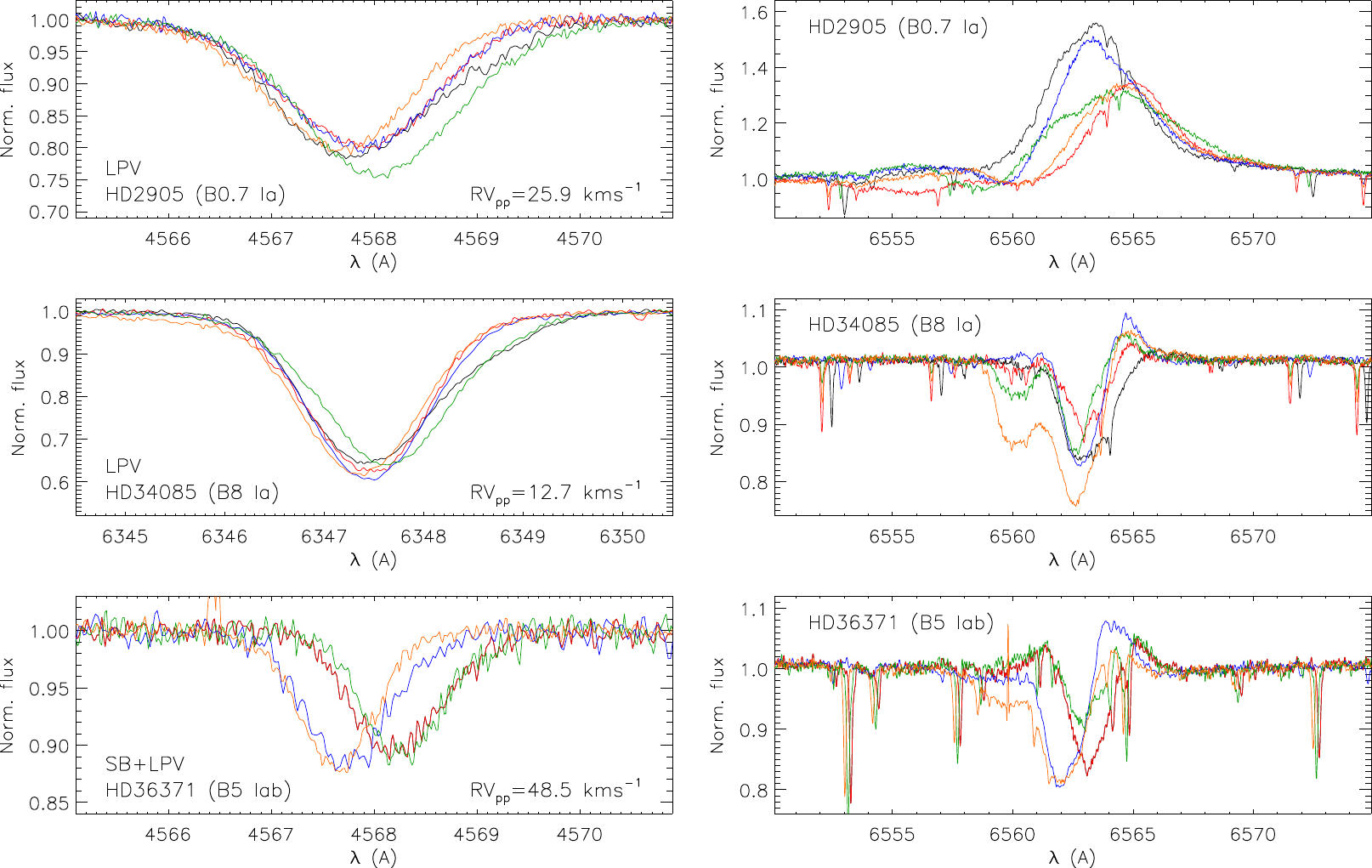}
\caption{Three illustrative examples of spectroscopic variability found in O and B-Sgs. The top and middle panels depict two stars in which the line-profile variability (LPV) detected in the diagnostic lines considered in this work to obtain RV estimates (left panels) is mainly produced by stellar oscillations. The bottom panel corresponds to a star in which the detected RV shifts are clearly associated with the orbital motion in a spectroscopic binary system (SB). We also show the corresponding spectroscopic variability detected in the H$\alpha$ line, which is mainly coupled with variability of the stellar wind.}
\label{figure_lpv_profs}
\end{figure*}

Figure~\ref{rv_cm1} shows the comparison of RV measurements obtained by means of both methodologies. For illustrative purposes, we include three clear spectroscopic binaries with a dominant component (i.e., detected as SB1 systems after a visual inspection of the behavior of the line profiles, see, for example, the bottom left panel in  Fig.~\ref{figure_lpv_profs}), and another three in which the detected variability is intrinsic, originated by stellar oscillations (see top and middle left panels in Fig.~\ref{figure_lpv_profs}).

Both techniques provide similar results, with an agreement better than 1~\kms\ in most cases. Additionally, as expected, larger discrepancies (but still $\lesssim$,2~\kms) are found in those cases where intrinsic variability gives rise to time-dependent asymmetric line profiles. This is exemplified by HD\,2905 and HD\,34085 (see Fig.~\ref{figure_lpv_profs}), where the aforementioned spectral characteristic leaves a clearer imprint in the RV estimate obtained by the centroid technique than when considering Gaussian fitting.

We note that these time-dependent asymmetries can be produced not only by stellar oscillations but also by the presence of spots on the stellar surface \citep[e.g.,][]{Aerts2014}, variable stellar wind emission, and/or the presence of a much dimmer secondary component in a binary system. In the latter case, even if the lines of the secondary star may not be easily detectable by visual inspection of the individual spectra, they are expected to produce a temporal variation of the skewness of the line profiles of the more luminous star in anti-phase with the radial velocity resulting from Gaussian fitting. Consequently, in all these cases, the radial velocity estimates associated with the centroid of the line profile are expected to be more affected by non-binary effects than the Gaussian fitting. Therefore, for the purposes of this study, we considered the radial velocity measurements from Gaussian fitting more appropriate. Additionally, for better interpretation of results, we visually inspected the detected variability in the position and shape of the line profiles of the diagnostic line used to obtain the radial velocity estimates, as well as the H$\alpha$ line (see some examples in Fig.~\ref{figure_lpv_profs}).

\begin{figure}[!t]
\centering
\includegraphics[angle=90, width=0.47\textwidth]{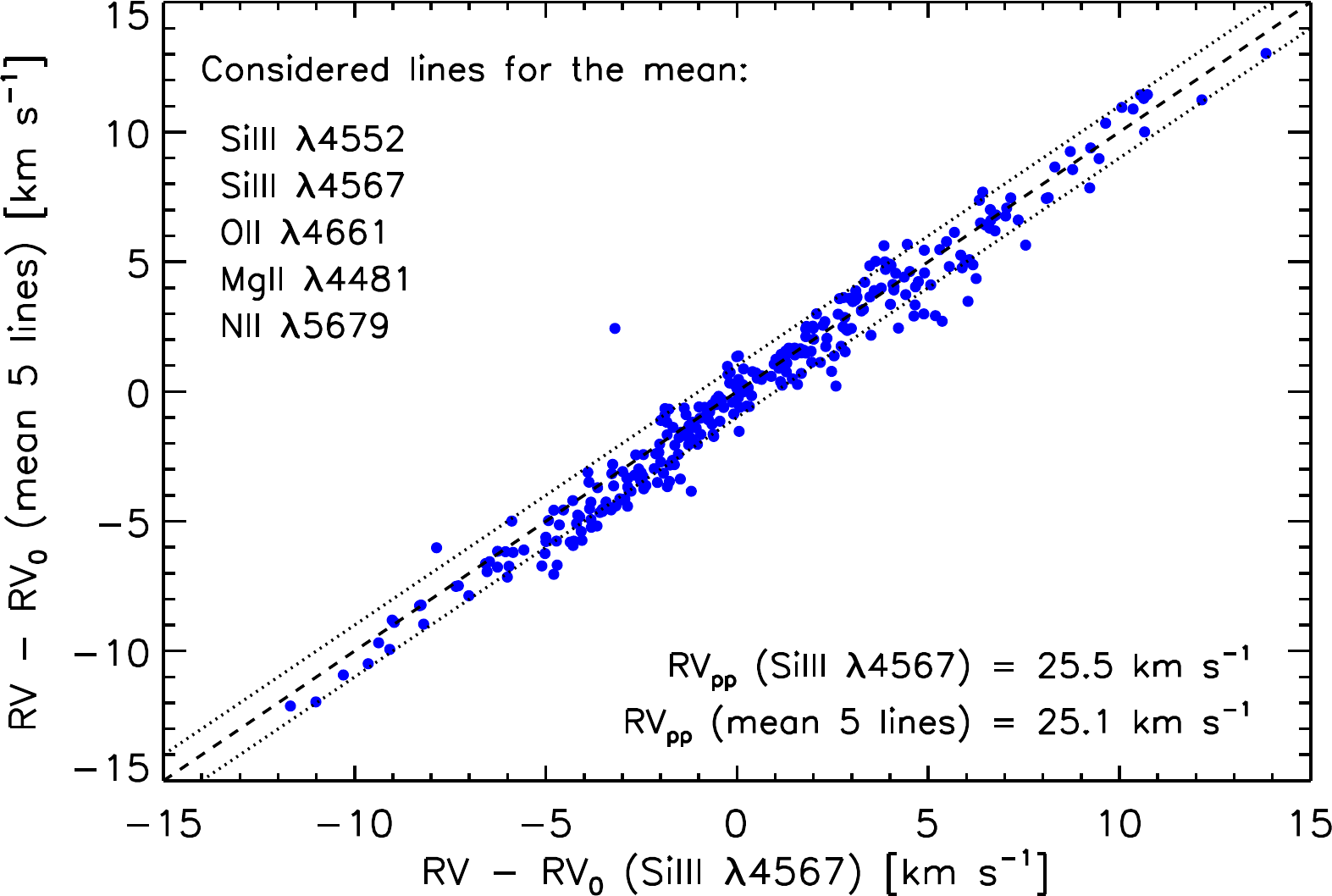}
\caption{Comparison of two different sets of RV measurements in the early B-Sg HD~2905. On the abscissa, those based on only one diagnostic line (Si\,{\sc iii}$\lambda$4567); on the ordinate, those corresponding to the mean RV per epoch computed from individual measurements obtained for the 5 different lines quoted in the plot. Diagonal dashed and dotted lines represent the 1:1 relation and the $\pm$1~\kms\ deviation, respectively.}
\label{figure_rvs}
\end{figure}

\begin{figure*}[!t]
\centering
\includegraphics[angle=90, width=0.98\textwidth]{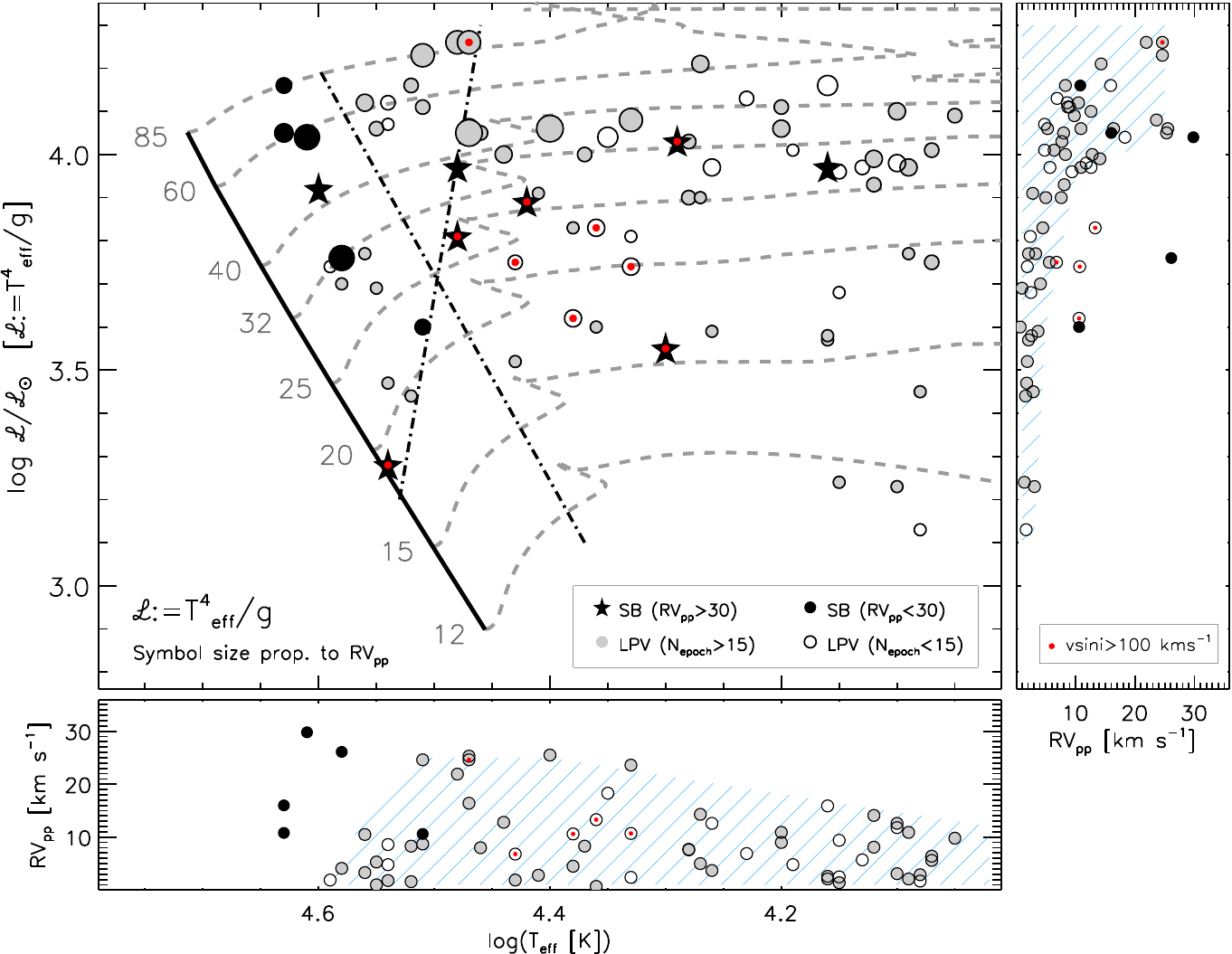}
\caption{(Top left) Spectroscopic HR diagram for the studied sample of stars. Symbol sizes are proportional to the measured peak-to-peak amplitude of RV variability, where those stars having \rvpp\,>\,30~\kms\ are represented with a star symbol. Open white and grey circles indicate stars for which the detected RV variability is attributed to stellar oscillations. Black circles and star symbols correspond to clearly detected spectroscopic binaries. Grey circles indicate stars for which we have more than 15 epochs, while red dots designate stars with \vsini\,>\,100~\kms. Evolutionary tracks (and the location of the Zero Age Main Sequence, ZAMS, line) are from non-rotating models by \cite{Ekstrom2012}. The almost vertical dashed-dotted line roughly separates the O and B star domains. The inclined dashed-dotted line parallel to the ZAMS roughly separates dwarf/subgiant stars from the more evolved giants/supergiants. (Top right and bottom left) Detected \rvpp\ as a function of log~$\mathcal{L}$ and log~\Teff, respectively. The symbols have the same meaning as in the left panel (note that stars with \rvpp\,>\,30~\kms\ are outside the limits of each panel). The blue striped areas indicate the parameter space in which the detected line-profile variability can be explained in terms of stellar oscillations.}
\label{fig-hr}
\end{figure*}

We also conducted a formal exercise to investigate the possibility of minimizing the effect of stellar oscillations on the RV measurements by combining results obtained from lines of different ions and equivalent widths. The outcome of this investigation is summarized in Fig.~\ref{figure_rvs}, where we present a comparison of results for six different diagnostic lines for the case of the early B-supergiant pulsating star HD\,2905 \citep{SimonDiaz2018}.

We found that the combined use of different lines did not help eliminate the effect of stellar oscillations or significantly improve the uncertainty associated with individual RV measurements. As illustrated in Fig.~\ref{figure_rvs}, all measurements are consistent with each other, without systematic differences across all epochs. Therefore, we decided to base our final estimates on only one diagnostic line, namely the one also used to determine the line-broadening parameters (see Sect.~\ref{section31}).

\section{Results}\label{section4}

Following the arguments presented in Sect.~\ref{section32}, we use the peak-to-peak amplitude of the radial velocity distribution (\rvpp) resulting from the Gaussian fit of a specific diagnostic line to quantify the detected variability of our program stars.

Table \ref{table:B1} summarizes the total number of considered spectra, as well as the final values of RV$_{\rm pp}$ for each star in the sample. The table is organized by bins of increasing \rvpp\ and, within each bin, by spectral type, ranging from early O-dwarfs to late B-Sgs. Column 4 indicates the diagnostic line used for the line-broadening (Sect.\ref{section31}) and RV (Sect.\ref{section32}) analysis. Uncertainties associated with the measurement of RVs are approximately 1\,--\,3~\kms\ (depending on the line-broadening), except for a few stars with large projected rotational velocities (\vsini~$\ge$~200 \kms), where the uncertainties may reach up to 5\,--\,10~\kms.

The top left panel in Fig.~\ref{fig-hr} provides a global overview of the results of our study in the spectroscopic HR diagram. The size of the symbol associated with each star is proportional to the measured \rvpp\ (except for those targets where \rvpp\ is larger than 30~\kms, which are highlighted with a star symbol). Further notes about the various symbols and colors used in this figure can be found in the corresponding caption.

This panel is complemented by two additional panels presenting the distribution of detected \rvpp\ as a function of \Teff\ (bottom panel) and log~$\mathcal{L}$ (right panel). As described in Sect.~\ref{section5}, the utility of these two latter panels is twofold. On the one hand, they illustrate the difficulties imposed by the ubiquitous presence of spectroscopic variability due to stellar oscillations in the O and B supergiant domain, complicating the clear identification of single line spectroscopic binaries based solely on the \rvpp\ quantity. On the other hand, they provide the first statistically significant empirical map of the characteristic amplitude of radial velocity variability associated with pulsational phenomena in the entire O star and B supergiant domain.

We note that, although we have included a few O-type dwarfs and subgiants (13) in our study for comparison purposes regarding the detected variability associated with stellar oscillations, our primary focus henceforth will be on the results obtained for the more evolved OB-Sgs (56) and early B-Gs (5). For a detailed summary of results from a more in-depth study of spectroscopic binarity among Galactic O-type stars, we refer the reader to \citet{Barba2017} and references therein.

\begin{figure}[!t]
\centering
\includegraphics[angle=90, width=0.48\textwidth]{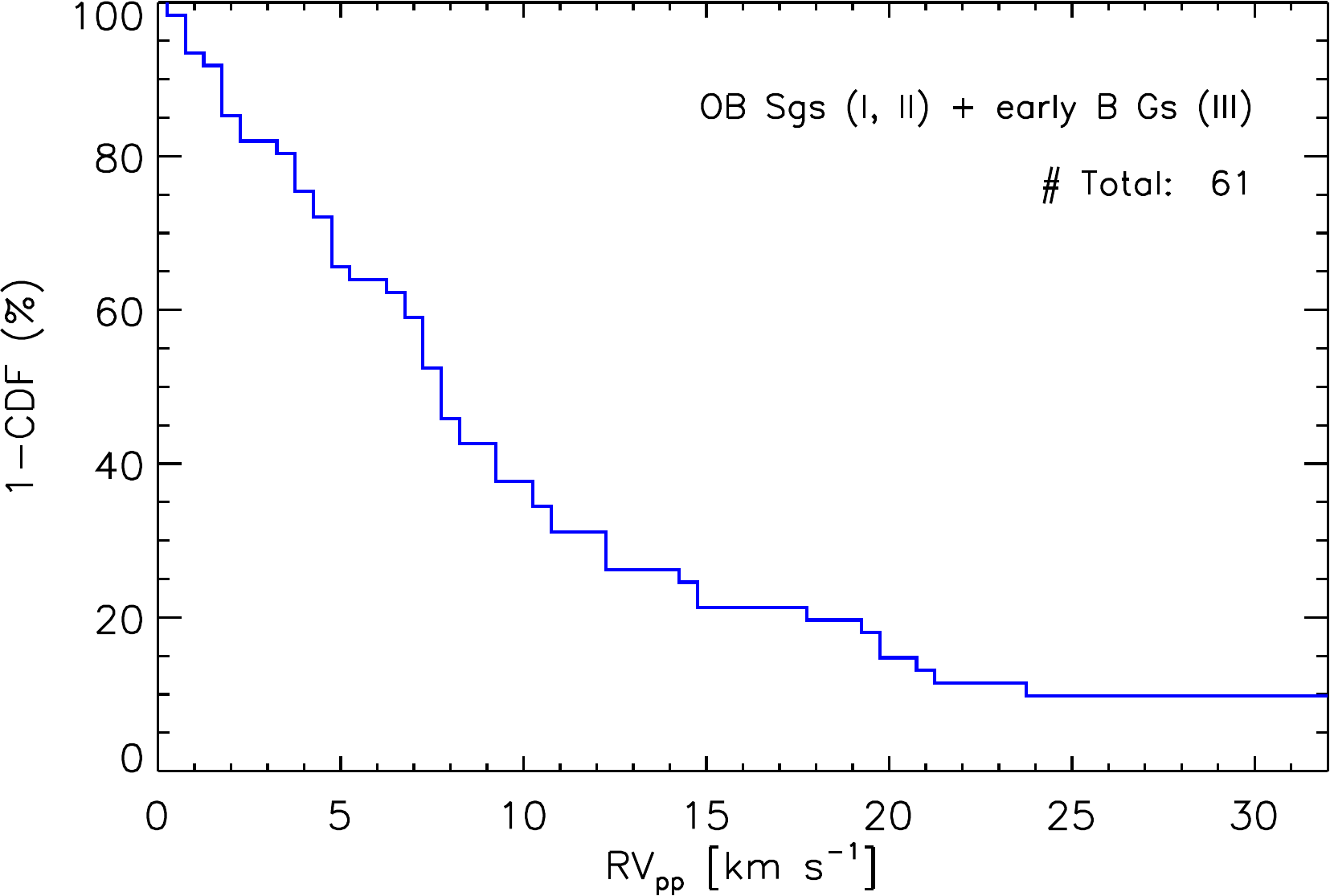}
\caption{Percentage distribution of the 56 OB~Sgs (plus five late-O/early-B giants) as a function of the measured \rvpp. CDF: cumulative distribution function.}
\label{figure_bins}
\end{figure}

In this regard, Fig.~\ref{figure_bins} depicts the percentage distribution of this specific sample of evolved high mass stars as a function of the measured value of \rvpp. In this case, we assume a bin size of 0.5~\kms\ and cut the distribution at $\sim$30~\kms, leaving out five early B-Sgs (of luminosity class II) and one mid B-Sg with peak-to-peak amplitudes of the radial velocity curves ranging from $\sim$40 to $210$~\kms\ (see top left panel in Fig.~\ref{fig-hr} and Table~\ref{table:B1}). 

Lastly, for completeness, we summarize in Table~\ref{table.summarySB2} the percentage of SB2 systems detected by visual inspection of all spectra available in the IACOB spectroscopic database at the time of performing this study (see Sect.~\ref{section2}). We focus particularly on three groups of stars of interest for this study: first, the late O- and early/mid/late B-type supergiants (OB-Sgs), which are the main stellar group discussed hereafter. For comparative purposes, we also consider the O-type dwarfs and giants in the sample as one group\footnote{This information has been extracted from Table E.2 in \cite{Holgado2020}.}, along with the early B-type giants as the third group.

In summary, the percentage of SB2 systems is much smaller among the OB-Sg sample compared to their less evolved counterparts, the O-type dwarfs/giants, and the less luminous early B-Gs. Notably, no SB2 systems with a mid- or late-B~Sg component were found. This result will be further discussed, along with the percentage of detected SB1 systems in the OB-Sg domain in Sect.~\ref{section53}.

\begin{table}
  \centering
  \caption[]{Summary of detected SB2 systems among three different massive star groups, based on the inspection of spectra in the complete sample of O- and B-type stars surveyed by the IACOB project.}
    \label{table.summarySB2}
      \begin{tabular}{lcccr}
\hline
\hline
\noalign{\smallskip}
        & Covered SpT \& LC & N$_{\rm stars}$  & \multicolumn{2}{c}{N$_{\rm SB2}$} \\  %
\noalign{\smallskip}
\hline
\noalign{\smallskip}
OB-Sgs      & O8\,--\,B9~I\,--\,II & 162  & 10  & 6$\pm$2~\% \\ %
O-Dws/Gs & O4\,--\,O9.7~V\,--\,III & 290  & 92  & 32$\pm$3~\%  \\ %
early B-Gs  & B0\,--\,B3~III & 37   & 9   & 24$\pm$8~\%  \\ %
\hline
      \end{tabular}
\end{table}

\section{Discussion}\label{section5}


In light of the stability of the FIES and HERMES spectrographs, which is better than approximately 1 \kms, one could reasonably consider tagging all stars in the sample with a measured \rvpp\,$\gtrsim$\,2\,--\,3\,\kms\ as single line spectroscopic binaries. Under this assumption, the percentage of detected SB1 stars in our sample of 61 OB-Sgs and late-O/early-B giants would be approximately 90\,\%.

However, the situation is more complex, as one must consider several other sources of spectroscopic variability present in the blue supergiant domain, including pulsations and/or wind-variability phenomena. See \citet{SimonDiaz2010, SimonDiaz2018, Aerts2017, Aerts2018, Burssens2020} for further discussion and references. 
For example, this is illustrated in Fig.~\ref{figure_bins} \citep[see also][]{Dunstall2015}, where we demonstrate that the final percentage of stars among our sample of OB-Sgs which can be definitively identified as SB1 critically depends on the chosen threshold in \rvpp\ used to minimize the number of false positives arising from intrinsic sources of variability. For instance, assuming a threshold in \rvpp\ for clear spectroscopic binary detection of 5~\kms\ would result in a percentage of around 65\%; however, this percentage drops to approximately 15\% if the threshold is set at 20~\kms. Therefore, before drawing any firm conclusions, we need to better understand the impact of intrinsic stellar variability in the OB-Sg domain.

\subsection{The impact of intrinsic variability when hunting for SB1 systems in the OB-Sg domain}\label{section52}

Previous investigations of spectroscopic binaries in O-type stars and B-Sgs have mostly relied on no more than 5\,--\,25 spectra distributed over several months. The extremely poor duty cycle of the compiled observations hinders the possibility of performing a proper assessment of the characteristic amplitude of RV variability due to factors such as stellar oscillations. Consequently, the separation of clear spectroscopic binaries from presumably single pulsating stars has been based only on indirect and not very robust arguments \citep[e.g.,][]{Sana2013, Dunstall2015}.

While in recent years, the number of studies incorporating larger multi-epoch observational datasets (with much higher cadence) for the investigation of spectroscopic variability in O stars and B-Sgs has slowly but surely increased\footnote{\cite[see also previous works by][]{Fullerton1996, Prinja1996, Kaufer1997, Rivinius1997, Markova2005, Markova2008}.} \citep[e.g.,][]{Moravveji2012a, Kraus2015, Martins2015, SimonDiaz2017, SimonDiaz2018, Aerts2017, Aerts2018, Hauke2018}, we still lack a complete overview of the characteristic amplitude of RV variations produced by stellar oscillations in the entire OB-Sg domain.

\begin{figure*}[!t]
\centering
\includegraphics[angle=90, width=0.98\textwidth]{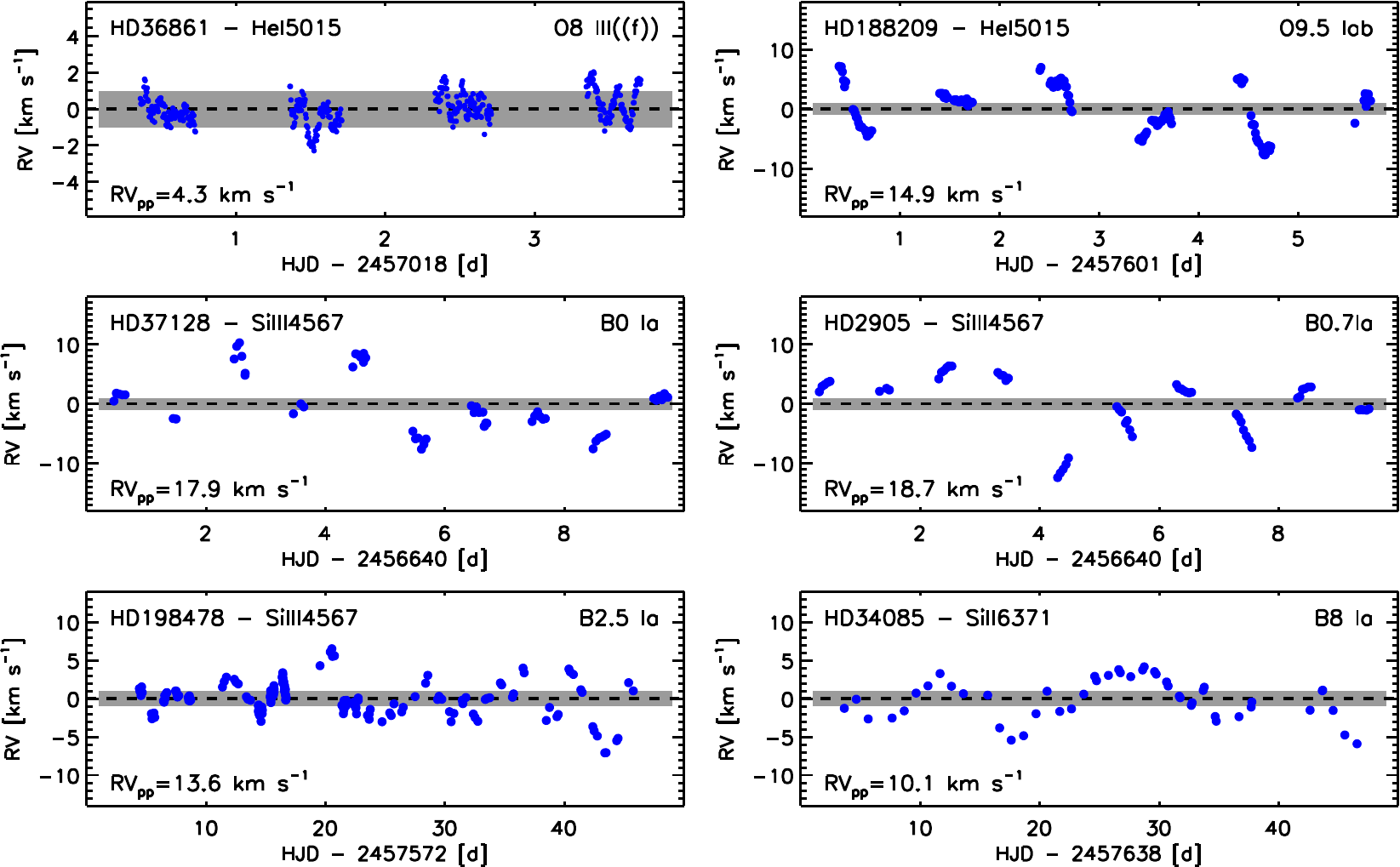}
\caption{Example of RV curves obtained with the {\em Hertzsprung}-SONG telescope in a sample of presumably single stars showing intrinsic variability originating from stellar oscillations. Grey horizontal bands indicate, in all panels for comparison purposes, the $\pm$1~\kms\ range.}
\label{figure_rv_curves}
\end{figure*}

This paper addresses this limitation thanks to the large number of targets (and the associated coverage of the spectroscopic HR diagram) for which we have more than 30\,-\,40 epochs. Indeed, to better illustrate the importance of being aware of the effect of pulsational-type phenomena when hunting for single-line spectroscopic binaries among massive OB stars, we decided to make use of some additional long-term and high-cadence spectroscopic observations obtained with the Hertzsprung-SONG telescope of a selected sample of bright northern O stars and B-Sgs (see Sect.~\ref{section2}).

Figure~\ref{figure_rv_curves} shows examples of the type of RV variability detected in several presumably single pulsating OB-Gs and Sgs observed with the SONG telescope. The selected stars cover a range in spectral type from O8 to B8, and the time-span and cadence of the compiled observations differ from one target to another, as indicated by the labels on the horizontal axis. In particular, these observational characteristics range from a few days (time-span) and minutes (cadence) in the case of the O8\,III((f)) star HD\,36861 ($\lambda$\,Ori\,A, Meissa) to several months and one day for the late BSg HD\,34085 (Rigel), respectively. In all cases, the detected temporal behavior of the measured RV is clearly not associated with the orbital motion in a binary system but rather with intrinsic variability of the star, likely associated with different types of stellar pulsations (e.g., p- and g-modes driven by the $\kappa$- or the $\varepsilon$-mechanisms), convectively driven internal gravity waves, and/or wind variability.

It is out of the scope of this paper to describe in detail the origin of the detected spectroscopic variability of the six stars presented in Fig.~\ref{figure_rv_curves} \citep[see, however,][]{Moravveji2012b, Kraus2015, Aerts2017, SimonDiaz2017, SimonDiaz2018, Burssens2020}. However, these illustrative examples can serve as a guide in the identification of situations in which RV variation due to intrinsic variability could be misinterpreted as a signature of orbital motion in a spectroscopic binary system.

There are many other stars among the list of targets investigated in this paper (and depicted in Fig.~\ref{fig-hr}) for which we have enough FIES and HERMES spectra\footnote{See, e.g., those stars in Table~\ref{table:B1} for which we have more than 15 spectra.} to ascertain with a high degree of confidence that the measured \rvpp\ is more likely produced by intrinsic variability than by orbital motion in binary systems. We refer the reader to Fig.~\ref{figure_lpv_profs} for two illustrative examples of the type of line-profile variability detected in the case of HD\,2905 and HD\,34085, associated with the RV curves depicted in middle and bottom right panels of Fig.~\ref{figure_rv_curves}, respectively.

By using this first set of non-SB1 stars (tagged as line-profile variables -- LPV -- in the last column of Table~\ref{table:B1} and highlighted with grey circles in Fig.~\ref{fig-hr}), we were able to provide, for the first time, some empirical hints about the parameter space in the spectroscopic HR diagram in which the detected line-profile variability can be (empirically) explained in terms of intrinsic stellar variability (see blue striped areas in Fig.~\ref{fig-hr}). To a first order, one can consider that any star located within the blue striped area in both the \rvpp\ vs. \Teff\ and \rvpp\ vs. log~$\mathcal{L}$ diagrams cannot be confidently labelled as SB1 and must be consequently tagged (at least provisionally) as LPV. Conversely, if the star is located outside one of the two blue striped areas (or both), it is quite likely a single-line spectroscopic binary (except for those cases with a \vsini\ larger than $\sim$100~\kms, which must be investigated separately, since in these cases the boundaries of the blue striped areas are naturally extended to larger values of \rvpp\ due to the broader line-profiles \citep[see those cases marked with a red dot in Fig.~\ref{fig-hr}, and also more specific examples in the case of fast rotating O-type stars in][]{Britavskiy2023}).

\begin{figure}[!t]
\centering
\includegraphics[angle=90,width=0.49\textwidth]{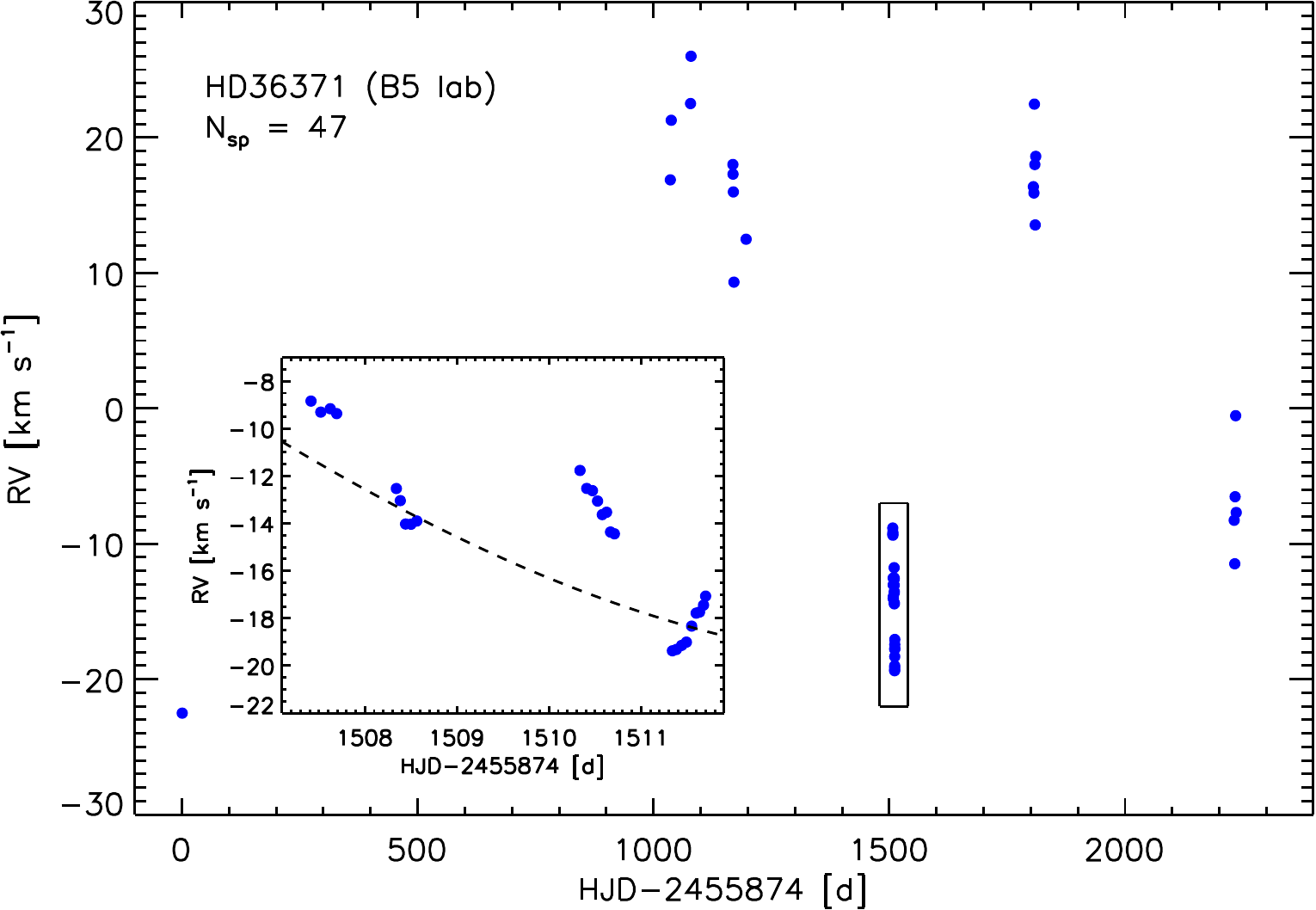}
\caption{Example of a mid B-Sg with RV variations resulting from its combined binary and pulsating nature. The inset shows how the typical intrinsic variability affecting the single mid B-Sgs, which is of the order of 10\,--\,15~\kms, is producing non-negligible departures in RV from the orbital motion of the detected component in this SB1 system (highlighted with the dashed curve).}
\label{figure_lpv}
\end{figure}

Using the information presented in Figs.~\ref{fig-hr} and \ref{figure_rv_curves}, plus other available RV curves, not presented in the paper, we can extract the following conclusions:
\begin{itemize}
    \item Intrinsic variability due to stellar oscillations in the O and B-Sg domain can lead to values of \rvpp\ of the order of $\sim$20\,--\,25\,\kms\ in some diagnostic lines. Hence, one must be careful when tagging as SB1 stars showing RV variations below that threshold.
    \item The amplitude of intrinsic variability depends on the stellar parameter (or spectral class) domain. This variability is very small for O-type dwarfs, then increases towards the late-O and early-B~Sgs and decreases again towards the mid- and late-B~Sgs. Also, the detected intrinsic variability increases with luminosity of the star for a given \Teff\ (or SpT). Therefore, a different threshold in \rvpp\ for spectroscopic binary detection should be considered for stars in different regions of the sHRD. 
    \item The typical time-scales of intrinsic variability in the O and B-Sg domain increase as a star with a certain mass increases its size (see, e.g. Fig.~\ref{figure_rv_curves}, where the size of the star is directly proportional to the spectral type for the considered targets). 
    \item All these characteristics of the detected intrinsic variability in O stars and B-Sgs will have an impact in the required strategy to be able to efficiently disentangle the orbital motion in a binary system from other sources of spectroscopic variability. Indeed, one must keep in mind that, as illustrated in Fig.~\ref{figure_lpv}, both effects are acting together in a binary system, leaving a combined imprint in the radial velocity curve. 
    \item Consequently, single-line spectroscopic binaries in which the \rvpp\ originated by the orbital motion is of the order or smaller than the intrinsic variability of the brighter companion are going to be very difficult to detect spectroscopically. Therefore, we might be missing some low amplitude SB1 systems.
\end{itemize}


\subsection{Percentage of spectroscopic binaries in the OB-Sg domain}\label{section53}

Taking into account the information described in Sect.~\ref{section52}, as well as our previous expertise about the type of line-profile variability due to stellar pulsations commonly found in O stars and B~Sgs \citep[see, e.g., Fig. 10 in][]{SimonDiaz2017}, we proceeded with a careful evaluation (on a star-by-star basis) of the potential spectroscopic binary status of all stars in our initial working sample (which also includes the 13 O-type dwarfs/subgiants).  To achieve this, we conducted an individual visual inspection of each radial velocity curve, paying attention to the specific location of the star in the various panels of Fig.~\ref{fig-hr}, as well as to the behavior of the diagnostic line used for the RV measurement and the H$\alpha$ line.

Stars marked with a black symbol in Fig.~\ref{fig-hr} correspond to those cases in which we confidently identify the target as a clear SB1. These mainly correspond to cases in which the large number of available epochs allowed us to find RV points that deviate from the detected intrinsic variability, or stars with peak-to-peak variations in RV larger than 40~\kms.  
As mentioned in the previous section, the threshold in \rvpp\ associated with a clear SB1 detection is smaller in the case of O-type dwarfs/subgiants (around 10\,--\,15~\kms) than in the case of OB-Sgs and early B-Gs (reaching values up to $\sim$25~\kms). In total, we have identified four O-type dwarfs, three late-O/early~B-Sgs, two early~B-Gs, and one mid~B-Sg. All of them are marked as SB1 in the last column of Table~\ref{table:B1}. 

This result implies a minimum percentage of single line spectroscopic binaries in the OB~Sg domain of 10$\pm$4\% (or 13$\pm$5\% spectroscopic binaries of any type if we also take into account the two early-B~Sg SB2 systems identified in the complete IACOB sample). Alternatively, if we only account for the stars with luminosity class I and/or II, the corresponding percentages are 7$\pm$3\% and 11$\pm$4\%, respectively.

\subsection{Comparison with the literature}\label{section54}

There are few previous studies in the literature investigating the incidence of binarity among B-Sgs. One of the most notable studies, \citep{Dunstall2015}, was conducted as part of the VLT-FLAMES Tarantula Survey (VFTS; \citep{Evans2005}).

\cite{Dunstall2015} employed a cross-correlation method to estimate \rvpp\ in a sample of 408 B-type stars in the 30 Doradus (30 Dor) region of the Large Magellanic Cloud (LMC). Among them, 47 were classified as supergiants\footnote{Defined as having a logarithm gravity, \grav$<$3.3 dex. We note that this definition exclude the O-Sgs and the early B-Gs.}. Notably, two of these were previously identified as SB2 systems by \cite{Howarth2015}. By employing a threshold of 16~\kms\ to distinguish intrinsic variability from clear binary detection, they identified a total of 9 SB1 systems (with an additional 4 targets having \rvpp\ in the range 12\,--\,15~\kms, considered as likely SB1s). Including the 2 detected SB2 systems, this led to a binary fraction of 23$\pm$6\% (or 30$\pm$7\% if likely SB1s are also included). 

Interestingly, the percentage of B-Sgs found in 30~Dor is very similar to the one that we would infer in our study (see Fig.~\ref{figure_bins}) by assuming a threshold of 16~\kms\ (alternatively 12~\kms). However, as we have demonstrated in Sect.~\ref{section53}, the number of clearly confirmed SB1s could be slightly overestimated in \cite{Dunstall2015}, since some (if not all) of the B-Sgs labeled as SB1, but having a \rvpp\ in the range 12\,--\,25~\kms, may actually be spurious detections of spectroscopic binaries. Indeed, by inspection of the bottom panel of Fig.~2 in \citeauthor{Dunstall2015}, one can estimate that the percentage of B-Sgs with \rvpp$>$25~\kms\ drops down to $\sim$17\%. 

Another important outcome of the VLT-FLAMES Tarantula survey refers to the establishment of the binary fraction among the O-type stars in 30~Dor. We remind that the B-Sgs are expected to be the direct descendants of these O-type stars; therefore, the comparison of results obtained in both stellar domain should offer us direct empirical constraints about massive binary evolution. \cite{Sana2013} reported a spectroscopic binary fraction of 35$\pm$3\% among the sample of 360 O-type stars surveyed by VFTS (or 51$\pm$4\% when they correct for observational biases). In this case, they considered as single line spectroscopic binaries those targets displaying statistically significant radial velocity variations with an amplitude of at least 20~\kms. This result is in fairly good agreement with other independent surveys of O-type stars in the Milky Way finding that the percentage of binaries with at least one O-type component range between 40 and 70\%. 

Therefore, there is increasing empirical evidence that many O-type binaries do not reach the BSg phase as clear spectroscopic binaries \citep[see also results presented in][indicating that the binary percentage decreases as massive O-type stars evolve along the Main Sequence]{Barba2017}. However, we must interpret the exact numbers published in the literature with caution, as different studies may employ different strategies to separate actual spectroscopic binaries from likely single intrinsically variable stars. For example, while \cite{Sana2013} assumed a 20~\kms\ threshold to identify clear binaries, our observations suggest that the intrinsic spectroscopic variability of O giants is expected to be lower, typically below 10~\kms. Unlike the situation in \cite{Dunstall2015} for B-Sgs, \cite{Sana2013} may be underestimating the percentage of O-type spectroscopic binaries in 30~Dor, particularly among dwarfs and giants.

In summary, we emphasize once more that identifying spectroscopic binaries among O-type stars and B-Sgs cannot rely solely on establishing a single threshold to separate genuine binaries from intrinsically variable isolated stars. The effectiveness of such thresholds depends on the location of the star in the HRD. Furthermore, it is crucial to consider that stars identified as LPVs could potentially be small-amplitude and/or long-period binaries, where the signature of orbital motion is obscured within the RV variability produced by stellar pulsations.

\section{Summary and conclusions}\label{summary}

Thanks to the intensive observational efforts of several groups over the last two decades, we are gradually gaining a comprehensive empirical understanding of spectroscopic binarity in the early evolutionary stages of stars with masses exceeding approximately 15 to 20 solar masses. This includes both Main Sequence O-type stars and blue supergiants.

Our study reveals a significant finding: there is mounting evidence indicating a substantial decrease, by a factor $\sim$4\,--\,5, in the incidence of detected spectroscopic binaries when transitioning from O-type stars to the domain of blue supergiants (B-Sgs). This trend is evident in both our Galactic B-Sg sample and the LMC study by \cite{Dunstall2015}. Notably, there is a notable scarcity of spectroscopic B-Sg binaries with mid and late spectral types. Furthermore, the occurrence of SB2 systems among B-Sgs is minimal, comprising less than 5\% of the total, in contrast to the higher incidence observed among O-type dwarfs, which have been reported to reach levels as high as 40\% \citep{Barba2017}.

In qualitative terms, a significant portion of the identified double-line spectroscopic binaries, particularly those with at least one O-type star component, are anticipated to transition to single-line (SB1) systems as the more massive component evolves. This transformation can occur under various scenarios: (a) the initially more massive O-type star evolves into a more luminous B-Sg without substantial interaction with its lower-mass companion, (b) it sheds enough mass through mass transfer to become a fainter, stripped lower-mass star, or (c) it survives a supernova explosion as a neutron star or black hole, with the binary system remaining intact. In each of these cases, the contrast in luminosity between the two binary components increases, making the detection of clear SB2 systems more challenging and providing a natural explanation for the scarcity of SB2 systems in the B-Sg domain.

Given the argument presented above, one might naively anticipate a rise in the proportion of detected SB1 systems among B-Sgs compared to their less evolved O-star counterparts. However, this assertion appears to contradict empirical observations. Yet, we must consider that the situation is more intricate from both an evolutionary and observational standpoint.

On one hand, the majority of close massive binaries originate from either case A or case B mass transfer and typically manifest as binaries with a B-type supergiant component only intermittently in later stages of evolution \citep[see, e.g.,][]{Wang2020}. Moreover, a considerable fraction of these binaries may merge during main sequence evolution, estimated to be around 10\% \citep{deMink2014}. Additionally, some B-Sgs have been hypothesized to be in a post-red supergiant phase of evolution, either evolving as single stars \citep[e.g.,][]{Ekstrom2012} or resulting from a merger between a red supergiant and a lower mass main sequence companion \citep[e.g.,][]{Menon2019, Menon2024}.

On the other hand, wider binaries may evolve to a scenario where the O-type component transitions into a mid/late B-Sg without filling its Roche lobe. These systems typically have long orbital periods and consequently exhibit relatively low-amplitude RV curves. As a result, intrinsic variability due to stellar oscillations, which, as discussed in Sect.~\ref{section52}, can induce RV variations of up to 20\,--\,25\,\kms, can hinder the clear identification of orbital motion in these systems.

While more work is still needed to reach firm conclusions, it becomes evident that investigating the incidence of binarity in the B-Sg domain can provide valuable empirical insights into several open questions in the field of massive stars. These include the percentage of merger events occurring during the main sequence evolution of O-type stars \citep{deMink2014}, the exact location of the Terminal Age Main Sequence \citep{McEvoy2015}, the potential post-red supergiant evolutionary nature of B-Sgs \citep{deMink2014}, and ultimately, the impact of binaries on massive star evolution \citep{Langer2012, Wang2020}, including intermediate phases in which an OB star has a black hole and/or neutron star companion \citep{Langer2020}.

In this regard, we want to emphasize once more the need for caution when combining and interpreting the information provided by various available studies. This caution is necessary not only due to potential observational biases affecting the samples being considered but also because of the impact that intrinsic variability has on the detection of actual spectroscopic binaries, especially among OB-Sgs.

Indeed, the second important message we want to highlight from our study is that defining the exact percentage of spectroscopic binaries in the OB-Sg domain requires a better understanding and characterization of the amplitude of spectroscopic variability produced by the various sources of intrinsic variability predicted to occur in these types of stars.

Crucially, we remark that depending on the origin of this intrinsic variability, there may be significant dependencies on metallicity. Hence, any empirical investigation performed in the Milky Way (like the one presented in this paper) might not necessarily be directly extrapolated to other metallicity environments. For example, at present, there is an intense debate about the main physical driver of the detected stellar variability (both photometric and spectroscopic) and the so-called macroturbulent broadening in the entire massive OB star domain \citep[see, e.g.,][and references therein]{SimonDiaz2017, Grassitelli2015a, Bowman2020}. Depending on the relative impact that the convectively driven internal gravity waves predicted to originate in the stellar core \citep{AertsRogers2015} and/or the subsurface convective layers \citep{Cantiello2009, Grassitelli2015a, Grassitelli2015b, Cantiello2019, Lecoanet2019} can have on the stochastic low-frequency variability, which is increasingly detected in blue massive OB stars \citep{Bloome2011, Buysschaert2015, Aerts2017, Aerts2018, SimonDiaz2018, Bowman2019, Bowman2020}, stellar metallicity may play an important role.

This is, therefore, a crucial research avenue that will undoubtedly shape our future understanding of the prevalence of binary systems throughout the evolution of massive stars. In this pursuit, it is imperative to consider other anticipated sources of stellar variability, such as coherent heat-driven modes \citep{Godart2017}, and intrinsic variations in the stellar wind. As exemplified by \cite{Burssens2020}, the TESS mission presents a unique opportunity for advancements in this field.

\begin{acknowledgements}
Based on observations made with the Nordic Optical Telescope, operated by NOTSA, and the Mercator Telescope, operated by the Flemish Community, both at the Observatorio de El Roque de los Muchachos (La Palma, Spain) of the Instituto de Astrof\'isica de Canarias. In addition, this paper makes use of observations obtained with the Hertzsprung SONG telescope operated on the Spanish Observatorio del Teide on the island of Tenerife by the Aarhus and Copenhagen Universities and by the Instituto de Astrof\'isica de Canarias. S.S-D, N.B., G.H and A.dB acknowledge support from the Spanish Ministry of Science and Innovation (MICINN) through the Spanish State Research Agency through grants PGC-2018-0913741-B-C22, PID2021-122397NB-C21, and the Severo Ochoa Programe 2020-2023 (CEX2019-000920-S). 
This work has also received financial support from the Canarian Agency for Economy, Knowledge, and Employment and the European Regional Development Fund (ERDF/EU), under grant with reference ProID2020010016.
N.B. also acknowledges support from the postdoctoral program (IPD-STEMA) of Liege University. N.C. acknowledges support from the Deutsche Forschungsgemeinschaft (DFG, German Research Foundation) - CA2551/1-1. G.H. acknowledges that this research has been partially funded by Spanish Ministry of Science and Innovation MCIN(10.13039/501100011033) through grant PGC2018-95049-B-C22 and 2022AEP005 and by "ERDF A way of making Europe". NC acknowledges funding from the Deutsche Forschungs-gemeinschaft (DFG) - CA 2551/1-1,2. 
\end{acknowledgements}


\begin{appendix} 

\clearpage
\onecolumn

\section{Tables and extra figures}\label{appendixB}

\begin{longtable}{lrrlcrccccl}
\caption{List of stars considered in this study separated by ranges of detected variability in radial velocity} \label{table:B1} \\
\hline
Star & SpC & N$_{\rm sp }$ & Line & RV$_{\rm pp}$ & $\Delta$T & log~\Teff & log $\mathcal{L}$/$\mathcal{L}_{\odot}$ & $v$~sin$i$ & $v_{\rm mac}$  &  Notes \\
&  &  & &  [\kms] & [d] &  [\Teff\ in K] & [dex] & [\kms] & [\kms]  &   \\ 
\hline
\endfirsthead
\multicolumn{11}{c}%
{{\bfseries \tablename\ \thetable{} -- continued from previous page}} \\
\hline
Star & SpC & N$_{\rm sp }$ & Line & RV$_{\rm pp}$ & $\Delta$T & log~\Teff & log $\mathcal{L}$/$\mathcal{L}_{\odot}$ & $v$~sin$i$ & $v_{\rm mac}$  &  Notes \\
&  &  & &  [\kms] & [d] &  [\Teff\ in K] & [dex] & [\kms] & [\kms]  &   \\ 
\hline
\endhead
\hline \multicolumn{11}{r}{{Continued on next page}} \\
\endfoot
\hline
\endlastfoot
%
\multicolumn{11}{l}{RV$_{\rm pp}~\leq$~5 \kms} \\
\hline
   HD\,47839 &   O7\,V((f))z &  109 &   O\,{\sc iii} &	3.4 & 3210 & 4.58 & 3.70 &  43 &  67 & LPV$^{b}$  \\  
  HD\,164492 &  	   O7.5\,Vz &  13  &   O\,{\sc iii} &	2.3 & 2552 & 4.59 & 3.74 &  39 &  54 & LPV? \\  
   HD\,46966 &  	   O8.5\,IV &  47  &   O\,{\sc iii} &	3.6 & 1502 & 4.56 & 3.77 &  40 &  66 & LPV  \\  
  HD\,214680 &  	      O9\,V &  38  &   O\,{\sc iii} &	0.9 & 3579 & 4.55 & 3.69 &  14 &  43 & LPV? \\  
   HD\,36512 &  	    O9.7\,V &  22  &   O\,{\sc iii} &	1.0 & 3578 & 4.52 & 3.44 &  13 &  33 & LPV? \\  
   HD\,34078 &  	    O9.5\,V &  32  &   O\,{\sc iii} &	1.1 & 3576 & 4.54 & 3.47 &  13 &  32 & LPV? \\  
   HD\,36861 &         O8\,III((f)) &  866 &   O\,{\sc iii} &	4.1 & 3332 & 4.55 & 4.06 &  52 &  77 & LPV \\  
  HD\,218376 &  	  B0.5\,III &  19  &  Si\,{\sc iii} &	1.3 & 1938 & 4.43 & 3.52 &  31 &  54 & LPV \\  
  HD\,205139 &  	     B1\,Ib &  24  &  Si\,{\sc iii} &	1.9 & 1127 & 4.41 & 3.91 &  36 &  63 & LPV \\  
  HD\,119608 &  	     B1\,Ib &  10  &  Si\,{\sc iii} &	2.4 &  749 & 4.33 & 3.81 &  45 &  71 & LPV \\  
   HD\,24398 &  	     B1\,Ib &  38  &  Si\,{\sc iii} &	4.2 &  729 & 4.38 & 3.83 &  45 &  59 & LPV  \\  
   HD\,52089 &  	   B1.5\,II &  20  &  Si\,{\sc iii} &	0.7 &	59 & 4.36 & 3.60 &  25 &  50 & LPV? \\  
   HD\,31327 &  	   B2.5\,Ib &  30  &  Si\,{\sc iii} &	3.5 & 2239 & 4.26 & 3.59 &  34 &  53 & LPV \\  
   HD\,14134 &  	     B3\,Ia &  12  &  Si\,{\sc iii} &	4.7 & 1550 & 4.19 & 4.01 &  38 &  58 & LPV \\  
   HD\,24432 &  	     B3\,II &  9   &   Si\,{\sc ii} &	2.0 &  400 & 4.15 & 3.68 &  28 &  49 & LPV \\  
  HD\,164353 &  	     B5\,Ib &  32  &   Si\,{\sc ii} &	2.2 & 1870 & 4.16 & 3.57 &  28 &  52 & LPV \\  
  HD\,191243 &  	     B5\,Ib &  30  &   Si\,{\sc ii} &	2.5 & 2449 & 4.16 & 3.58 &  27 &  40 & LPV \\  
  HD\,175156 &  	     B5\,II &  21  &   Si\,{\sc ii} &	1.4 &  687 & 4.15 & 3.24 &  21 &  22 & LPV \\  
   HD\,14542 &  	     B8\,Ia &  7   &   Si\,{\sc ii} &  4.6 &  817 & 4.10 & 3.98 &  39 &  46 & LPV \\  
   HD\,12301 &  	     B8\,Ib &  31  &   Si\,{\sc ii} &	2.9 & 1159 & 4.08 & 3.45 &  32 &  41 & LPV \\  
   HD\,14322 &  	     B8\,Ib &  10  &   Si\,{\sc ii} &	4.4 &  817 & 4.07 & 4.01 &  38 &  46 & LPV \\  
   HD\,35600 &  	     B9\,Ib &  9   &   Si\,{\sc ii} &	1.2 &  817 & 4.08 & 3.13 &  30 &  32 & LPV? \\  
  HD\,212593 &  	    B9\,Iab &  28  &   Si\,{\sc ii} &	2.2 & 1810 & 4.09 & 3.77 &  25 &  33 & LPV \\  
\hline                                   
\multicolumn{11}{l}{5~$<$~RV$_{\rm pp}$~$\leq$~10 \kms} \\
\hline 
  HD\,192639 &  	 O7.5\,Iabf &    13 &	O\,{\sc iii} &   8.6 & 3292 & 4.54 & 4.12 &  82 &  95 & LPV \\   
   HD\,34656 &  	O7.5\,II(f) &    10 &	O\,{\sc iii} &   8.4 & 1870 & 4.56 & 4.12 &  67 &  77 & LPV \\   
  HD\,162978 &  	O8\,II((f)) &    10 &	O\,{\sc iii} &   5.1 & 2625 & 4.54 & 4.07 &  54 &  86 & LPV \\   
  HD\,207198 &    O8.5\,II((f)) &    70 &	O\,{\sc iii} &   8.1 & 2450 & 4.52 & 4.16 &  52 &  97 & LPV \\   
  HD\,209975 &  	     O9\,Ib &    51 &	O\,{\sc iii} &   7.6 & 1873 & 4.51 & 4.11 &  50 & 104 & LPV \\   
  HD\,204172 &  	     B0\,Ib &    19 &  Si\,{\sc iii} &   7.6 & 1808 & 4.46 & 4.05 &  68 &  85 & LPV \\   
     HD\,191877 &  	     B1\,Ib &     9 &  He\,{\sc i}   &   8.0 & 2542 & 4.33 & 3.74 & 162 &  95 & LPV \\  
   HD\,40111 &  	     B1\,Ib &    13 &  Si\,{\sc iii} &   7.3 & 2782 & 4.43 & 3.75 & 141 &  58 & LPV \\   
   HD\,91316 &  	    B1\,Iab &   636 &  Si\,{\sc iii} &   7.4 & 2849 & 4.37 & 4.00 &  50 &  70 & LPV \\   
  HD\,206165 &  	     B2\,Ib &    79 &  Si\,{\sc iii} &   7.6 & 2539 & 4.28 & 4.03 &  44 &  59 & LPV \\   
   HD\,14818 &  	     B2\,Ia &    17 &  Si\,{\sc iii} &   7.7 & 1451 & 4.28 & 3.90 &  43 &  70 & LPV \\   
  HD\,198478 &  	   B2.5\,Ia &    57 &  Si\,{\sc iii} &   8.8 & 1869 & 4.20 & 4.11 &  39 &  54 & LPV \\   
   HD\,42087 &  	   B2.5\,Ib &    31 &  Si\,{\sc iii} & 	 5.0 & 1200 & 4.27 & 3.90 &  43 &  56 & LPV \\  
   HD\,53138 &  	    B3\,Iab &    12 &  Si\,{\sc iii} &   6.7 & 2233 & 4.23 & 4.13 &  37 &  56 & LPV \\   
  HD\,225094 &  	    B3\,Iab &    21 &  Si\,{\sc iii} &   9.9 & 1864 & 4.20 & 4.06 &  35 &  57 & LPV \\   
   HD\,13267 &  	     B5\,Ia &    12 &	Si\,{\sc ii} &   9.6 &  816 & 4.15 & 3.96 &  34 &  57 & LPV \\   
    HD\,7902 &  	     B6\,Ib &    65 &	Si\,{\sc ii} &   8.4 &  999 & 4.12 & 3.93 &  38 &  55 & LPV \\   
   HD\,46769 &  	     B8\,Ib &    24 &	Si\,{\sc ii} &   5.2 &  341 & 4.10 & 3.23 &  70 &  23 & LPV \\   
  HD\,208501 &  	     B8\,Ib &    14 &	Si\,{\sc ii} &   5.7 & 1867 & 4.13 & 3.97 &  41 &  52 & LPV \\   
  HD\,202850 &  	    B9\,Iab &    32 &	Si\,{\sc ii} &   5.4 & 1957 & 4.07 & 3.75 &  41 &  37 & LPV \\   
   HD\,21291 &  	     B9\,Ia &    40 &	Si\,{\sc ii} &   9.8 & 1810 & 4.05 & 4.09 &  41 &  29 & LPV \\   
\hline  
\multicolumn{11}{l}{10~$<$~RV$_{\rm pp}$~$\leq$~15 \kms} \\
\hline  
  HD\,46223 &  	 O4\,V((f)) &  6   &   C\,{\sc iv} &	10.8 & 1871 & 4.63 & 4.16 &  51 & 112 & SB1?  \\    
  HD\,202214 &  	   O9.5\,IV &  25  &  Si\,{\sc iii} &  11.6 & 1866 & 4.51 & 3.60 &  26 &  36 & SB1? \\  
  HD\,188439 & 	     B0.5\,IIIn &  8   &  He\,{\sc i}   &  12.9 &  783 & 4.38 & 3.79 & 310 &  ... & LPV \\
   HD\,38771 &  	   B0.5\,Ia &  245 &  Si\,{\sc iii} &  14.7 & 3331 & 4.47 & 4.06 &  53 &  83 & LPV \\  
  HD\,213087 &  	   B0.5\,Ib &  28  &  Si\,{\sc iii} &  12.8 & 1869 & 4.44 & 4.00 &  61 &  91 & LPV \\  
   HD\,41117 &  	     B2\,Ia &  34  &  Si\,{\sc iii} &  10.7 & 1938 & 4.27 & 4.21 &  38 &  67 & LPV \\  
   HD\,14143 &  	     B2\,Ia &  14  &  Si\,{\sc iii} &  12.5 & 1897 & 4.26 & 3.97 &  50 &  69 & LPV \\  
    HD\,4841 &  	     B5\,Ia &  10  &   Si\,{\sc ii} &  11.3 &  316 & 4.12 & 3.99 &  40 &  59 & LPV \\  
   HD\,34085 &  	     B8\,Ia &  123 &   Si\,{\sc ii} &  12.7 & 2538 & 4.10 & 4.10 &  37 &  51 & LPV \\  
  HD\,199478 &  	     B8\,Ia &  27  &   Si\,{\sc ii} &  11.0 & 1867 & 4.09 & 3.97 &  42 &  46 & LPV \\  
\hline                                             
\multicolumn{11}{l}{15~$<$~RV$_{\rm pp}$~$\leq$~20 \kms} \\
\hline  
  HD\,164794 & 	    O4\,V((f))z &  9   &	C\,{\sc iv} &  16.0 & 2627 & 4.63 & 4.05 &  62 &  95 & SB1?\\
  HD\,188209 &  	  O9.5\,Iab &  122 &   O\,{\sc iii} &  15.4 & 2379 & 4.48 & 4.26 &  54 &  93 & LPV\\
   HD\,13854 &  	    B1\,Iab &  14  &  Si\,{\sc iii} &  18.2 & 1552 & 4.35 & 4.04 &  53 &  67 & LPV \\
  HD\,190603 &  	  B1.5\,Ia+ &  68  &  Si\,{\sc iii} &  19.9 & 3606 & 4.33 & 4.08 &  45 &  67 & LPV \\
   HD\,25914 &  	     B6\,Ia &  10  &   Si\,{\sc ii} &  15.4 & 1435 & 4.16 & 4.16 &  38 &  33 & LPV \\
\hline                                   
\multicolumn{11}{l}{20~$<$~RV$_{\rm pp}$~$\leq$~30 \kms} \\
\hline  
   HD\,46150 &      O5\,V((f))z &  18  &	C\,{\sc iv} &  29.8 & 2902 & 4.61 & 4.04 &  60 & 112 & SB1 \\
   HD\,35619 &    O7.5\,V((f))z &  9   &   O\,{\sc iii} &  26.1 & 2034 & 4.58 & 3.76 &  40 &  60 & SB1 \\
  HD\,188001 &    O7.5\,Iabf &  98  &   O\,{\sc iii} &  20.0 & 2027 & 4.51 & 4.23 &  69 & 100 & LPV$^{c}$ \\
   HD\,30614 & 	      O9\,Ia &  86  &	O\,{\sc iii} &  24.0 & 3330 & 4.47 & 4.26 & 113 &  77 & LPV$^{c}$ \\
   HD\,37128 & 	      B0\,Ia &  274 &  Si\,{\sc iii} &  21.9 & 3331 & 4.47 & 4.05 &  55 &  85 & LPV \\
  HD\,184915 & 	  B0.5\,IIIn &  10  &  He\,{\sc i}   &  20.0 & 1147 & 4.36 & 3.83 & 285 &   ... & LPV \\
    HD\,2905 & 	    B0.7\,Ia &  291 &  Si\,{\sc iii} &  {\bf 25.9} & 3331 & 4.40 & 4.06 &  60 &  82 & LPV \\
\hline                                   
\multicolumn{11}{l}{RV$_{\rm pp}$~$>$~30 \kms} \\
\hline 
   HD\,199579 &    O6.5\,V((f))z &  184 &	O\,{\sc iii} &  82.1 & 3576 & 4.60 & 3.92 &  52 &  90 & SB1 \\
    HD\,37041 & 	   O9.5\,IVp &  17  &	O\,{\sc iii} & 213.8 & 3279 & 4.54 & 3.28 & 133 &  40 & SB1 \\
    HD\,36486 & 	 O9.5\,IINwk &  62  &	O\,{\sc iii} & 223.9 & 2297 & 4.48 & 3.97 & 100 &  94 & SB1 \\
   HD\,25639  & 	     B0\,III &  7   &  He\,{\sc i} &  59.2 & 2623 & 4.48 & 3.81 & 269 &  ... & SB1 \\
   HD\,187459 & 	    B0.5\,II &  7   &  Si\,{\sc iii} & 207.7 & 2542 & 4.42 & 3.89 & 150 &  76 & SB1 \\
   HD\,187879 & 	     B1\,III &  17  &  Si\,{\sc iii} & 186.6 & 1362 & 4.30 & 3.55 & 101 &  66 & SB1$^{d}$ \\
   HD\,47240  &           B1\,II &  11  &  Si\,{\sc iii} &  79.0 &	1074 & 4.29 & 4.03 & 105 &  74 & SB1 \\
   HD\,36371  & 	     B5\,Iab &  47  &	Si\,{\sc ii} &  48.5 & 2235 & 4.16 & 3.97 &  32 &  45 & SB1 \\
    \hline
    \multicolumn{11}{p{0.90\textwidth}}{
    \begin{list}{} 
\small{
\item[$^{a}$] The various diagnostic lines considered for the line broadening analysis and the measurement of RVs are O\,{\sc iii}~$\lambda$5591, C\,{\sc iv}~$\lambda$5811, Si\,{\sc iii}~$\lambda$4567, Si\,{\sc ii}~$\lambda$6347 and He\,{\sc i}~$\lambda$5875. This later line was only used in a few cases in which the other considered metal lines were too shallow due to the fast rotation or high effective temperature of the star.
\item[$^{b}$] HD\,47839 is a well known spectroscopic binary including a dominant O7~V component and a much fainter fast rotaing early-B dwarf; the period of this system is, however, of several years, and hence, given the time-span coverage of our observations, we do not detect and important contribution of the orbital motion to the RV variability in the investigated spectra.
\item[$^{c}$] HD\,30614 and HD~188001 are identified as spectroscopic binaries in SB9 \citep{Pourbaix2004}; however, despite we measure a \rvpp$\sim$20\,--\,25~\kms in both cases, our high cadence SONG observations (not shown in Fig.~\ref{figure_lpv}) clearly indicate that the detected spectroscopic variability is due to pulsations, and not binarity \citep[see also further notes in][]{Trigueros2021}.
\item[$^{d}$] Despite HD\,187879 has been sometimes identified as a SB2 system \citep[see, e.g.][]{Tkachenko2014}, we do not detect any signature of the companion in our spectra.
}
\end{list}
} \\
\end{longtable}

\clearpage
\twocolumn

\begin{figure*}
\centering
\includegraphics[angle=0, width=0.99\textwidth]{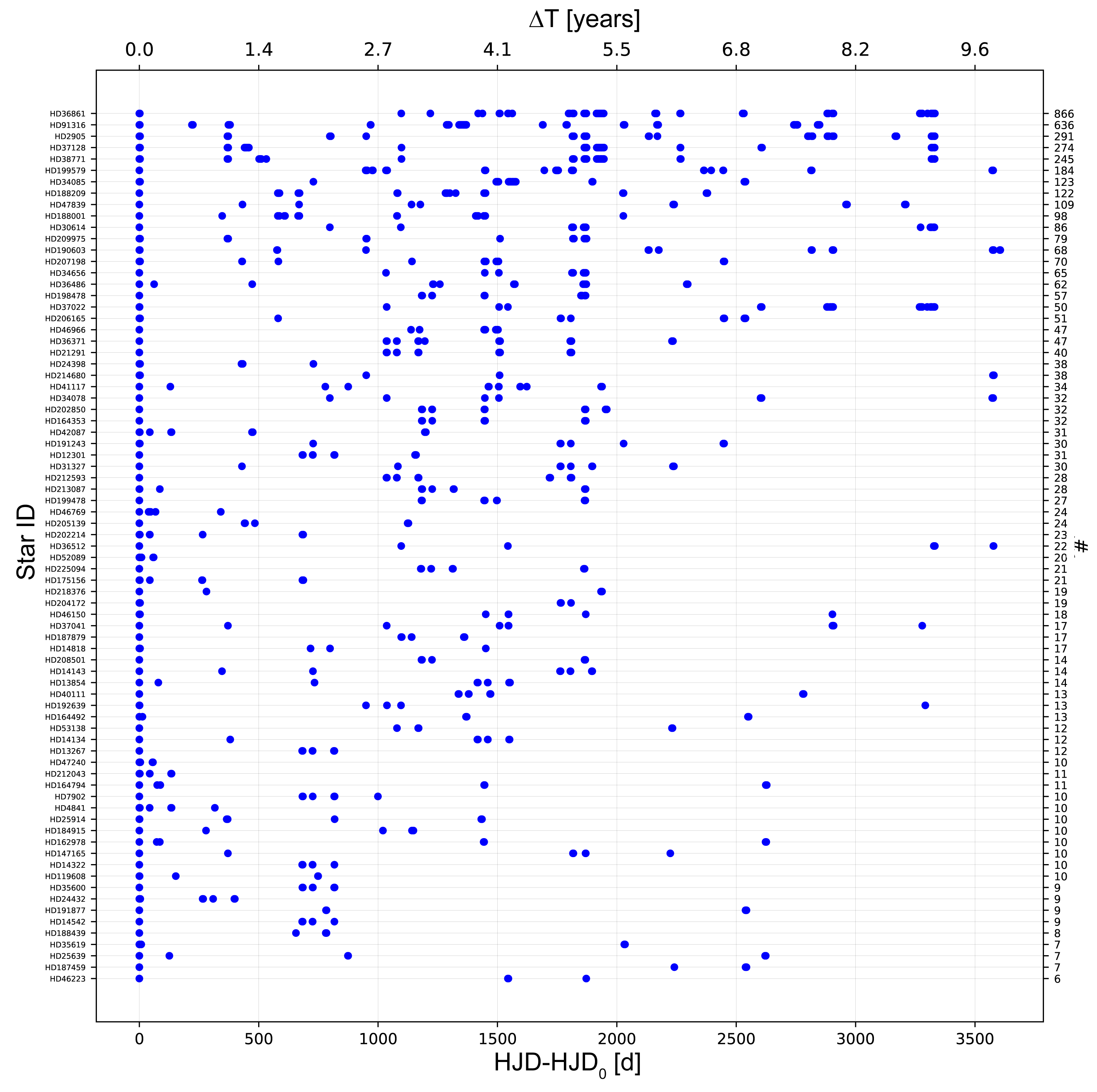}
\caption{General overview of the number of epochs, time~sampling and total time~span associated with the studied sample of O stars, B-Sgs and early~B-Gs not identified as SB2. HJD$_{\rm 0}$ refers to the time when the first observations for each individual target was obtained}.
\label{figure1a}
\end{figure*}

\end{appendix} 

\end{document}